\documentclass[a4paper]{article}

\usepackage[english]{babel}

\usepackage[a4paper,top=2cm,bottom=2cm,left=3cm,right=3cm,marginparwidth=1.75cm]{geometry}

\usepackage{amsmath}
\usepackage{graphicx}
\usepackage[colorlinks=true, allcolors=blue]{hyperref}
\usepackage{epstopdf}
\usepackage{txfonts}
\usepackage{authblk}

\newcommand*\andnewline{%
        \end{tabular}
        \begin{tabular}[t]{c}
}
\title{Paper Title}
\author{M.~Aoki\textsuperscript{1},
A.~M.~Baldini\textsuperscript{2},
R.~H.~Bernstein\textsuperscript{3},
C.~Carloganu\textsuperscript{4},
S.~Mihara\textsuperscript{5},
S.~Miscetti\textsuperscript{6},
T.~Mori\textsuperscript{7},
W.~Ootani\textsuperscript{7},
F.~Renga\textsuperscript{8},
S.~Ritt\textsuperscript{9} and
A.~Sch\"oning \textsuperscript{10} 
\andnewline
\textsuperscript{1}Osaka University, Osaka, Japan \andnewline
\textsuperscript{2}Istituto Nazionale di Fisica Nucleare - Sez. di Pisa \andnewline
\textsuperscript{3}Fermi National Accelerator Laboratory, Batavia, Illinois, USA \andnewline
\textsuperscript{4}Universit\'e Clermont Auvergne, CNRS/IN2P3, LPCA, F-63000 Clermont-Ferrand, France \andnewline
\textsuperscript{5}High Energy Accelerator Research Organization, Ibaraki, Japan \andnewline
\textsuperscript{6}Istituto Nazionale di Fisica Nucleare - Laboratori Nazionali di Frascati \andnewline
\textsuperscript{7}ICEPP, The University of Tokyo, 7-3-1 Hongo, Bunkyo-ku, Tokyo 113-0033, Japan \andnewline
\textsuperscript{8}Istituto Nazionale di Fisica Nucleare - Sez. di Roma \andnewline
\textsuperscript{9} Paul Scherrer Institute, Villigen, Switzerland \andnewline
\textsuperscript{10} Heidelberg University, Heidelberg, Germany
 \andnewline
  on behalf of the COMET, MEG, Mu2e, and Mu3e Collaborations\andnewline
}

\title{Charged Lepton Flavour Violations searches with muons: present and future}

\begin{document}
\maketitle

\begin{abstract}
Charged-lepton flavor violation (cLFV) is one of the most powerful probes for New Physics (NP). Since lepton flavor conservation is an accidental symmetry in the Standard Model (SM), it is naturally violated in many NP models, with contributions at the level of the current experimental sensitivities. Moreover, the negligible SM contributions would make the observation of cLFV unambiguous evidence of NP. It makes these searches extremely sensitive and, at the same time, extremely pure.

Thanks to the intense muon beams currently available, their intriguing upgrade programs, and the progress in the detection techniques, cLFV muon processes are the golden channels in this field. Experimental programs to search for $\mu^+ \to e^+ \gamma$, $\mu^+ \to e^+ e^+ e^-$ and the $\mu \to e$ conversion in the nuclear field are currently ongoing. We review the current status and the strategic plans for future searches.

This document is an update of the prior cLFV submission to the 2018 European Strategy for Particle Physics (ESPP); the earlier submission should be consulted for more experimental details~\cite{Baldini:2018uhj}.
\end{abstract}

\newpage

\section{Physics motivation}

The conservation of lepton flavor is an accidental symmetry in the Standard Model (SM), arising from its specific particle content (namely, the absence of right-handed neutrinos). As a consequence, this symmetry is typically lost in New Physics (NP) models, where lepton flavor violation is commonly predicted at the level of the current experimental sensitivities.
Indeed, the discovery of neutrino oscillations already demonstrated that this symmetry is not exact. Although neutrino oscillations are insufficient to give observable charged-lepton flavor violation (cLFV) effects, their existence further stimulates the search for this kind of process.

Since the late 1970s, intense muon beams have allowed us to search for cLFV with unparalleled sensitivity. The most simple cLFV muon decay is $\mu \to e \gamma$, for which the best limit on the branching ratio comes from the combination of the final MEG and the first MEG II results, $\mathcal{B}(\mu \to e \gamma) < 3.1 \times 10^{-13}$~\cite{MEGII-first}. It currently provides most of the strongest constraints on cLFV from NP models.
However, at tree level, searches for $\mu \to e \gamma$ are sensitive only to dipole-like lepton-photon vertices, while $\mu-e$ conversions and $\mu^+ \to e^+ e^- e^+$ are sensitive to both dipole and four fermion (contact) interactions. While the MEG II experiment is still taking data, new experiments to search for the other processes are currently under construction and are expected to provide their results within the next decade. The combination of results from the three cLFV modes will be critical to clarify the physics interpretation in the case of direct observations. If no positive signal is observed, extracting extensive constraints on the parameter space of NP models will also require limits from all these searches.

\section{Status of current experiments}

\subsection{MEG II}

The MEG II experiment~\cite{MEGII-design} is searching for the $\mu^+ \to e^+ \gamma$ decay at the Paul Scherrer Institut (PSI). The experiment is designed to reach a sensitivity of $6 \times 10^{-14}$ on the branching ratio of this decay, one order of magnitude below the sensitivity of the previous experiment MEG.

MEG II takes advantage of a high-intensity, continuous, positive muon beam delivered at PSI and stopped on a thin target in the $\pi$E5 area, with a nominal intensity of $4 \times 10^7$ stopped muons per second used for data taking. Continuous beams are necessary to search for rare muon decays, specifically $\mu \to e \gamma$, to minimize the rate of accidental coincidences of photons and positrons coming from different muon decays. This dominant background also limits the optimal beam current for the experiment: once the background reaches a non-negligible level, its rate scales with the beam intensity's square, making any further increase useless.

For a given beam intensity, the background level is determined by the experimental resolutions on the observables that allow us to detect the two-body kinematics of $\mu \to e \gamma$: the positron and photon energies (expected to be monochromatic at 52.8~MeV), their relative angle ($180^\circ$) and their time coincidence. 

In MEG II, the positron kinematics is reconstructed in a magnetic spectrometer instrumented with an extremely light, high-granularity cylindrical drift chamber and high-precision scintillating tiles for timing. A positron momentum resolution below 100~keV was reached, almost 4 times better than the performance of the MEG tracking system.
The innovative design of the drift chamber also inspired the concepts for gaseous inner trackers in future lepton colliders.

The photon is detected by an LXe calorimeter readout by PMTs and silicon photon detectors, which required the development of dedicated, UV-sensitive MPPCs. The detector provides measurements of energy ($\sim 2$~\% resolution), timing ($\sim 65$~ps resolution), and 
photon conversion position ($\sim 2.5$~mm resolution). The detector does not provide any measurement of the photon direction; that direction is inferred from the combination of the photon conversion point and the positron position at the target, assuming that both come from the same vertex.
The high granularity of the silicon photon detectors allowed MEG II to significantly improve the detector's performance with respect to MEG, particularly for photons converting near the entrance face of the calorimeter.

The experiment has been taking physics data since 2021 and is expected to run until the end of 2026, when the PSI beamlines will be shut down to be upgraded. The analysis of data collected in 2021 already provided a sensitivity competitive with the final one of MEG, and the combination of the two results provides the already mentioned best limit on the $\mu \to e \gamma$ branching ratio.

\subsection{Conversion experiments}

The search for the neutrinoless, coherent conversion of a muon to an electron in the field of a nucleus, $\mu^- N \to e^- N$, is carried out by the Mu2e experiment at FNAL and the COMET experiment at J-PARC. Both experiments share similar features and use an aluminum target to stop the muons. Their main goal is to increase sensitivity by four orders of magnitude compared to previous limits, allowing them to probe new physics mass scales of O($10^4~\mathrm{TeV}/c^2$).

The experimental signature of cLFV in conversion with aluminum is a monoenergetic electron with a momentum of 104.97~MeV/$c$. In both experiment designs, the signal should be well separated from the decay-in-orbit (DIO) electron spectrum, whose endpoint closely approaches the signal’s energy. Cosmic ray interactions in the apparatus can generate fake candidates, so a very efficient Cosmic Ray veto is required. A pulsed beam is used to reduce the prompt background coming from the Radiative Pion Capture (RPC) process, which is strongly suppressed by delaying the opening of the data acquisition gate by several hundred nanoseconds after the proton arrival times at the production target. Protons arriving out of sync with the main pulses can delay the RPC background and introduce a challenging source of background. Therefore, a high level of extinction ($>10^{10}$) of these protons is needed from the accelerator. Beyond the search for $\mu-e$ conversion, these experiments also allow us to explore other rare processes, such as muon-to-positron conversion, which violates lepton flavor and lepton number conservation\cite{1998334}. This process is a charged-lepton counterpart to neutrino-less double-beta decay, providing complementary insights into BSM physics. Additionally, searches for other relevant physics phenomena, such as axion-like particles (ALP) or dark photons, can be conducted by looking for bumps or deviations in the spectra of $\mu^- \to e^- X$\cite{PhysRevD.84.113010}, $\mu^+ \to e^+ X$\cite{PhysRev.181.1854, PhysRevD.91.052020, PhysRevD.101.052014}, and $\pi^+ \to e^+ X$\cite{PhysRevD.101.075002}. The latter two searches will require special running conditions, including changing the selected charge of the muon beam, reducing the magnetic field value in the detection volume, and adding material (degraders) to slow the pions so that they stop in a well-defined volume.

The construction of the experiments is nearly finished.  The timeline for the first data collection with beams and physics results aligns with the EU strategy.

\subsubsection{COMET}

\noindent
COMET aims to achieve the sensitivity at a level of $10^{-17}$ in the search for muon-to-electron conversion
in the field of an atomic nucleus ($^{27}$Al)\cite{comet-cdr}.
It takes a staged approach to reach this goal: COMET will start  the measurement with a shorter muon beam line
to demonstrate
the curved-solenoid technique along with physics data-taking (Phase-I\cite{comet-tdr}), and upgrade the facility to the final configuration
by incorporating improvements identified in Phase-I (Phase-II).
The target sensitivity of Phase-I is $3 \times 10^{-15}$ with an aluminum muon stopping target,
and is $3 \times 10^{-17}$ or better in Phase-II.
The Phase-I setup is currently under construction, with the goal of taking physics data starting in 2026.

COMET utilizes a high-quality 8-GeV pulsed proton beam produced by the J-PARC MR.
The time separation of the consecutive pulses is either $1.2\;\mu\mbox{s}$ or $1.8\;\mu\mbox{s}$.
The residual protons between the pulses have been measured to be smaller than $1.0 \times 10^{-10}$ (90\% C.L.)
for an instantaneous beam power equivalent to that of COMET Phase-I (3.2~kW)\cite{comet-extinction}.

The pulsed proton beam is extracted to the B-line of the J-PARC Hadron Hall.
The C-line branches off from the B-line to deliver the proton beam, impinging a production target in the Pion-Capture Superconductor Solenoid (PCS) located in the COMET experimental hall.
The PCS uses an aluminium-stabilized superconductor coil, 
which is 
lighter than a copper-stabilized coil. Thus, there is less nuclear heat load coming from the radiation flux from the production target.
The production target will be made of graphite for Phase-I,
and tungsten alloy for Phase-II.
Pions produced and captured by the strong magnetic field of the PCS
are transferred by the Muon Transport Solenoid (MTS) to the Detector Solenoid (DS).
The MTS is a curved solenoid with the capability to produce a vertical dipole field
that controls the vertical drift of charged particles.
Thanks to this dipole field, the MTS can 
select the charge and
tune the momentum of the muon beam and remove the contamination from higher momentum beam particles. 
The total bend angle of the MTS is $90^\circ$ for Phase-I
and will be extended to $180^\circ$ for Phase-II.

The DS will be located at the exit of the MTS in Phase-I.
Muons are stopped on a muon stopping target composed of aluminum disks located at the center of the DS.
The momentum of the electron from the muon decay will be measured with
the Cylindrical Drift Chamber (CDC)
with timing and particle type information provided by the Cylindrical Trigger Hodoscope (CTH).
The expected total number of muons stopped on the muon stopping target is $1.5 \times 10^{16}$ for Phase-I 
to achieve a sensitivity for the muon-to-electron conversion of $3 \times 10^{-15}$,
which is a two-order of magnitude improvement over SINDRUM-II.
Another objective in Phase-I is to confirm the function and performance of this PCS-MTS muon beam system. 
A stand-alone beam measurement with a dedicated detector system composed of straw-tube trackers and an electron calorimeter will also be conducted in Phase-I.
This 
will provide invaluable experience for the design of Phase-II.

In Phase-II, the solenoid magnet chain will be extended to
make better use of higher proton beam power. Another $90^\circ$ section of the MTS will be connected, forming a $180^\circ$ bend section to further reduce pion contamination in the muon beam.
A solenoid magnet with a graded magnetic field surrounding the muon stopping target will be included to increase the signal acceptance. Finally, another curved solenoid magnet also equipped with the momentum-selecting vertical magnetic fields will be inserted between the target and detector solenoid magnets to
prevent neutral and opposite-charge backgrounds from entering the detector region, as well as those which are not near the signal momentum range,
keeping the detector hit rate reasonable.
With these upgrades the proton beam power will be increased to 56~kW in Phase-II.

European contributions from France, Georgia, Germany, the UK, and JINR strengthen the COMET collaboration. Their contributions are fundamental to the success of the COMET project. They include 
the design and implementation of an online and offline software framework,
constructing a cosmic-ray veto, producing straw tubes for a tracker and LYSO crystals for a calorimeter, designing and fabricating the muon-stopping target, and more.  COMET European members strongly support the collaboration. 

The physics detector for Phase-I will be integrated
in early 2026, followed by commissioning with cosmic rays in the middle of 2026.
The first physics run will start in
2026 at low intensities and reduced radiation shielding.
The beam intensity will gradually increase as we add radiation shielding to reach the sensitivity goal of Phase-I.

\subsubsection{Mu2e}

Mu2e~\cite{Mu2e-tdr} is designed to improve the current limits on the ratio of conversions to nuclear capture by four orders of magnitude with a single event sensitivity of $3 \times 10^{-17}$. The data collection will take place in two phases with the same detector layout: Run-I, which will collect 10\% of the Mu2e full statistics before the FNAL long shutdown in 2028, and Run-II, which will complete the data collection starting in 2030.  Run-I will reach a sensitivity up to 1000 times better than the current limits and will also serve as a test phase to inform detector tuning/optimization, data reconstruction, and analysis strategies for Run-II. It will also contribute to preparing an advanced proposal for the design of the experiment upgrade (Mu2e-II), which will operate with the new beam from the PIP-2 linac, increasing statistics by another factor of 10 but with more demanding requirements for the detector, target design, and data handling. 

Mu2e uses an 8 GeV high-intensity proton beam, which is slowly extracted in pulses with a 1.7 us period. This beam hits a tungsten production target, producing pions that decay into muons.  A 23-meter-long solenoid system then directs about $1.5 \times 10^{-3}\mu/p$ muons to a $^{27}Al$ stopping target (the same material as COMET.) The solenoids are organized in three parts: a production solenoid (PS), where the pions are produced; a transport solenoid (TS), where pions decay to muons while being transported and charge selected; and a detector solenoid (DS), where the muon beam is delivered to the stopping target. Inside the DS, a high-precision detector operates within a 1 T magnetic field. The harsh experimental environment and the need for effective background suppression require innovative technologies, such as low-mass (15 $\mu$m thick) straw tube trackers, radiation-hardened electronics, and advanced calibration systems. Tracking ensures precise momentum measurement for the conversion energy of 105 MeV, $\sigma(\rm{p})< 180$ keV, while the undoped CsI crystal calorimeter with SiPM
readout provides timing $\sigma(\rm{t})< 500$ ps and energy information to improve background rejection. Additional challenges, such as minimizing muons interacting with residual gas, are addressed by operating in a vacuum environment at $10^{-4}$~Torr. The entire detector system resides inside the DS, which is surrounded by a cosmic ray veto system made of 300 m$^2$ of scintillation counters with SiPM readout, which provides $>$ 99.99\% cosmic ray rejection. A Stopping Target Monitor (STM) is used to identify muon capture X-rays and gamma lines with high-precision Germanium and LaBr$_3$ crystal detectors.

European researchers from Italy, UK, Germany, and JINR provide key contributions to the design, realization, and operation of sub-systems, such as the calorimeter and the STM, covering management roles and offering valuable expertise in data acquisition and analysis efforts. The TS magnet has already been installed and connected to the cryogenic system, and the PS magnet has been completed and will be delivered to the Mu2e hall in early Spring 2025. The DS cold mass has been completed, and the DS is in its final stages of construction. The detector systems~\cite{Mu2e-status} are nearing completion and will transition to operation in 2025, initially outside the DS, and later in 2026 within the DS. One specific capability of European researchers is the realization and usage of bent crystals for beam channeling, which will be applied for shadowing the beam in the slow extraction septa region. This technique will help reduce and control beam losses. 

\subsection{Mu3e}

The Mu3e experiment searches for the decay  $\mu^+ \to e^+ e^+ e^-$.
The search will be performed in two phases. The Mu3e Phase~I detector is designed for a muon stopping rate of $10^8$ per second, a rate that was already achieved at the Compact Muon Beamline (CMBL) in the $\pi$E5 area at PSI. 
To reach the ultimate sensitivity of  $B(\mu^+ \to e^+ e^+ e^-)<10^{-16}$, corresponding to an improvement of the existing limit \cite{SINDRUM:1987nra} by four orders of magnitude, muon stopping rates of $\approx 2 \cdot 10^9$ per second are required, which will become available at the High Intensity Muon Beamline  (HIMB) after 2028, see \autoref{sec:HIMB}. 
To cope with the higher background rates at HIMB, a new  detector (Mu3e Phase~II) with increased background suppression needs to be designed, see \autoref{sec:Mu3e2}.

The Mu3e Phase~I detector~\cite{Mu3e:2020gyw} consists of a four-layer pixel detector for particle tracking and vertexing, and two timing detector systems made from scintillating tiles and fibers, which have a time resolution of $\lesssim 100$~ps and $\approx 250$~ps, respectively. 
All detector data are zero-suppressed and continuously read out.  
Event reconstruction is performed online on a farm of CPUs. This includes reconstructing and fitting all tracks and 3-prong vertices, which are permanently stored for offline data analysis.
The experiment is performed in a helium atmosphere to minimize multiple Coulomb scattering inside a solenoid magnet providing a field of $B=1$~Tesla. Excellent momentum resolution of $\sigma(p) \ll 1$~MeV/c is achieved by a novel tracking detector design concept, which allows particles to be measured when they recurl in the magnetic field. 
The experiment includes several innovative technologies that have been specifically developed for Mu3e. This includes the Mupix High Voltage Monolithic Active Pixel Sensor (HVMAPS) for particle detection, the ultra-light tracking detector design with only $\approx 1.1 \cdot 10^{-3}$ radiation length per layer, the gaseous helium cooling system and the MuTrig ASIC, a multi-channel time-to-digital converter for SiPM signals.

Many components of the Mu3e Phase~I detector have been installed and are under commissioning.  A first physics data run is foreseen which includes all detectors of the central station in  2026. After the long shutdown of the HIPA accelerator in 2027 and 2028, data-taking will resume at the end of 2028 with all final detector components installed. 
For a running period of 300 days, a single event sensitivity of $\approx 2 \cdot 10^{-15}$ is expected. 
Parallel to the experimental activities, the Mu3e collaboration will continue R\&D for the Mu3e Phase~II experiment (\autoref{sec:Mu3e2}).

\section{Future experiments and facilities}



\subsection{HiMB at PSI} \label{sec:HIMB}

The High Intensity Proton Accelerator (HIPA) facility at PSI currently delivers the muon beam used for the MEG II and Mu3e Phase I experiments. In order to stop muons in a thin target, a low-energy, monochromatic beam is needed, which is achieved by selecting surface muons (i.e. muons produced by pions decaying at rest on the surface of the production target). In the $\pi$E5 area, were the experiments are performed, more than $10^8$ surface muons per second can be delivered.

Starting in 2027, the HIPA infrastructure at PSI will undergo a shutdown of about two years for an upgrade of the muon beam lines~\cite{Maso:2023zjp}. One of the two existing proton target stations will be dismantled and rebuilt. The thickness of the target and the slant angle with respect to the impinging proton beam have been optimized for the production of low-energy, positive muons. Two solenoids, oriented at $90^\circ$ with respect to the proton beam, will be used to capture secondary particles produced in the target. Two high-transmission beam lines, using exclusively solenoids for focusing, will transport the particles toward new experimental areas. Unlike quadrupoles, solenoids provide simultaneous focusing in both transverse directions, resulting in higher transport efficiencies.

Compared to the current highest-intensity, surface-muon beam lines at PSI, delivering up to $5 \times 10^{8}~\mu$/s, the new design will provide an increase of a more than a factor 4 in capture efficiency, and almost a factor 6 in transport efficiency, resulting in about a factor 25 higher intensity available in the experimental areas, which means more than $10^{10}~\mu$/s.

This remarkable advancement demands dedicated experimental initiatives to maximize its potential in uncovering forbidden muon decays~\cite{Aiba:2021bxe}.

\subsection{AMF at FNAL}

The Advanced Muon Facility (AMF) is a proposal for a next-generation muon facility at Fermilab that would exploit the full potential of the PIP-II accelerator to deliver the world’s most intense
$\mu^+$ and $\mu^-$ beams~\cite{corrodi2023workshopfuturemuonprogram}. This facility would enable broad muon science with unprecedented sensitivity, including a suite of CLFV experiments that could improve the sensitivity of planned experiments
by orders of magnitude, and study in detail the type of operators contributing to new physics in the case of an observation (e.g. high-$Z$ target in conversion experiments).
The AMF complex would use a fixed-field alternating gradient synchrotron (FFA) to create a
cold, intense muon beam with low momentum dispersion. Short, intense proton pulses are delivered
to a production target surrounded by a capture solenoid, followed by a transport system to inject
the muons produced by pion decays into the FFA ring. The phase rotation trades time spread for momentum spread, producing a cold, monochromatic muon beam. During that time ($\mathcal{O}$(1) $\mu$s), the pion contamination is reduced to negligible levels, and the FFA injection/extraction system effectively cuts off other sources of delayed and out-of-time backgrounds. The phase rotation requires a very short proton pulse, and a compressor ring is required to rebunch the PIP-II beam.

Muon-to-electron conversion experiments
require {\it negative} $\mu^-$ so a nucleus
can capture them. The decay experiments, 
$\mu^+ \rightarrow e^+ \gamma$ and $\mu^+ 
\rightarrow 3e$, require {\it positive} 
$\mu^+$, so the muons will not be captured in 
the material that slows and stops them.  
AMF's polarities can be reversed so that the 
storage ring holds $\mu^+$ instead of $\mu^-
$. Both conversion and decay experiments 
benefit from lowering the muon energy to 
decrease range straggling, the uncertainty in 
how deep into the stopping material the muon 
stops.  The FFA can be tuned to lower momenta 
than a stopped muon beam, or an induction 
linac following the FFA could be used to 
decelerate the beam. 

Finally, AMF is designed around a proton beam 
targeted inside a superconducting solenoid at 
$\mathcal {O}$ 500 kW -- 1MW.  The design 
could inform R\&D on the muon collider, which 
also uses a beam targeted inside a 
superconducting solenoid.  In that sense, AMF 
could be an R\&D step toward a muon collider.

\subsection{Mu2e-II}

An R\&D phase for Mu2e-II~\cite{Mu2e-II-snowmass} is also foreseen. EU researchers will lead the calorimeter R\&D, focusing on identifying better crystals and modifying electronics to enhance radiation hardness and improve the reconstruction quality of calorimetry, particularly when expecting higher occupancies in the detector. A similar path of R\&D will be pursued to explore alternatives for STM detectors and for new strategies for muon capture normalization. These R\&D efforts are currently under discussion and will become more focused following the startup of Run-I. Meanwhile, studies for new concepts of pion production targets \cite{Mu2e-II-target}, able to withstand the x10 intensity increase, are being carried out with a collaboration among Germany and US teams.

\subsection{Mu3e Phase-II} \label{sec:Mu3e2}

Building on the experience gained in constructing and operating the Phase 1 detector, Mu3e Phase 2 will profit from an improved experimental design and detector. 
For a factor of 20 increased muon stopping rate, as expected at HIMB, the accidental background will increase by a factor of 400 without any experimental changes.
The main background source is the time- and spatial coincidence of two ordinary muon decays, where both positron energies are close to the Michel edge (maximum $52.8$~MeV). 
Such events become $\mu^+ \rightarrow e^+ e^+ e^-$ candidates if both positrons have a back-to-back topology and one of the positrons undergoes Bhabha scattering, thus ``creating" a high energetic electron which is detected as third prong with the correct charge.
Multiple measures are foreseen to suppress this kind of background: the extension of the muon-stopping target, the reduction of the material in the stopping target region and the first tracking layer, as well as further improving the time and vertex resolution of HVMAPS. 

Currently, detector R\&D is carried out in several areas. This includes the construction of longer and thinner tracking detector modules, the development of HVMAPS with even smaller thickness and smaller pixel sizes for the inner tracking layer, as well as HVMAPS with $\approx 100$~ps time resolution to construct a dedicated timing layer, which will be able to stand the high particle rates. 
The experiment also plans to increase the magnetic field to 2 T (2.7 is possible) to improve the momentum resolution of the tracker.
Furthermore, improvements in services like detector readout (higher bandwidth) and serial powering (fewer cables) are foreseen to reduce the material budget.
Anticipating that detector R\&D will take until 2029 and a few years of construction time, data taking of Mu3e Phase~II is expected in the early 2030s. 
The ultimate sensitivity of $B(\mu^+ \rightarrow e^+ e^+ e^-) \approx 10^{-16}$, given by background from the radiative muon decays with internal conversion, will be reached after about three years of operation.

\subsection{Future \texorpdfstring{$\mu \to e \gamma$}\ ~experiments}

As already mentioned, the sensitivity of $\mu \to e \gamma$ searches is limited by the background coming from accidental coincidences of positrons and photons from different muon decays. Since this background scales with the square of the stopping muon rate, it prevents increasing the beam intensity above an optimal level, determined by the capability of discriminating the background itself at the trigger and data analysis level, which in turn comes from the detector resolutions. Moreover, the MEG II detector already suffers from pileup and aging effects due to the high beam rate. As a result, the optimal stopping rate for MEG II, around $4 \times 10^{7}~\mu/s$, is already more than a factor two below the maximum value that could be reached in the $\pi$E5 area at PSI. Without a significant improvement in the detector resolutions or a change of the experimental concept, there is no chance to exploit the much higher beam rates potentially available at HiMB or AMF.

In the last few years, a study group was initiated, with collaborators from MEG II and Mu3e, to develop a new concept for a $\mu \to e \gamma$ search exploiting beam rates up to $10^9 - 10^{10}~\mu$/s~\cite{Voena:2024vme}. The basic idea is to replace the calorimetric approach for photon detection with a conversion technique: thin layers of dense material are used to make the photons convert into $e^+e^-$ pairs, which are reconstructed by tracking detectors in a magnetic spectrometer. If the conversion material is an active one (e.g., a few cm of a very fast and highly luminous inorganic scintillating crystal), the combination with the spectrometer can give an energy resolution below 0.4~\%, a timing resolution below 40~ps, and a standalone measurement of the photon direction. The latter is not available in the MEG II calorimeter and will be extremely beneficial in reducing the accidental background by verifying the photon-positron vertex compatibility. Given the high susceptibility of gaseous trackers to pileup and aging processes at high rates, the positron could be reconstructed in a magnetic spectrometer instrumented with silicon detectors, in a design similar to the one of Mu3e (possibly with thinner sensors), in order to cope with extremely high occupancies at the price of a relatively small (or possibly null) loss in resolutions.

The low conversion probability for 52.8~MeV photons in a relatively thin conversion layer would limit the detection efficiency. However, it could be abundantly compensated by an increase in the beam intensity, while the better resolutions of the photon detector would permit a much better background discrimination. An improvement of one order of magnitude in sensitivity with respect to MEG II could be within reach~\cite{meg-next}. With an appropriate design, the positron detector could also be used for a next-generation $\mu \to e^+ e^+ e^-$ search after the completion of the Mu3e Phase-II program. 

The R\&D phase for the new photon and timing detectors can proceed alongside the other existing cLFV experimental programs, allowing for a first-phase experiment already during their operational lifetime. With a muon beam rate approaching $10^8~\mu$/s but still enabling the use of a conventional gaseous positron tracker, an improvement in sensitivity compared to MEG II could be already obtained.

A separate input document with additional details about this project is being submitted to the ESPP.

\section{Conclusions}

The efforts made in the last decades to develop an experimental program in the search for cLFV in the muon sector led to a family of top-class experiments which will deliver their results within this decade. In the meanwhile, upgrades of muon beam facilities are programmed or proposed, which could create the conditions for further, significant advancements. The unique sensitivity and cleanliness of new physics searches through cLFV in the muon sector call for an endeavor to develop new projects, with the aim of fully exploiting the new facilities and the expertise developed in these years. It will allow to maintain comparable sensitivities to all different processes and fully leverage their complementarity. 

\bibliographystyle{unsrt}
\bibliography{sample}

\end{document}